\documentclass[12pt,preprint]{aastex}

\usepackage[english]{babel}
\usepackage{natbib}

\shorttitle{Ethyl mercaptan}
\shortauthors{Kolesnikov\'{a} et al.}

\begin{document}

\title{Spectroscopic characterization and detection of Ethyl Mercaptan in Orion
\thanks{This work was based on observations carried out with the
IRAM 30-meter telescope. IRAM is supported by INSU/CNRS (France),
MPG (Germany) and IGN (Spain).}}

\author{L. Kolesnikov\'{a}}
\affil{Grupo de Espectroscop\'{i}a Molecular (GEM), Edificio Quifima, Laboratorios de Espectroscop\'{i}a y Bioespectroscop\'{i}a, Parque Cient\'{i}fico UVa, Unidad Asociada CSIC, Universidad de Valladolid, 47005 Valladolid, Spain.}
\email{lucie.kolesnikova@uva.es}

\and

\author{B. Tercero and J. Cernicharo}
\affil{Departamento de Astrof\'{i}sica, Centro de Astrobiolog\'{i}a CAB, CSIC-INTA, Ctra. de Torrej\'{o}n a
Ajalvir km 4, 28850 Madrid, Spain.}
\email{terceromb@cab.inta-csic.es, jcernicharo@cab.inta-csic.es}

\and

\author{J. L. Alonso and A. M. Daly}
\affil{Grupo de Espectroscop\'{i}a Molecular (GEM), Edificio Quifima, Laboratorios de Espectroscop\'{i}a y Bioespectroscop\'{i}a,
Parque Cient\'{i}fico UVa, Unidad Asociada CSIC, Universidad de Valladolid, 47005 Valladolid, Spain.}
\email{jlalonso@qf.uva.es, adammichael.daly@uva.es}

\and

\author{B. P. Gordon and S. T. Shipman}
\affil{Division of Natural Sciences, New College of Florida, Sarasota, FL 34243, USA.}
\email{brittany.gordon@ncf.edu, shipman@ncf.edu}

\begin{abstract}
New laboratory data of ethyl mercaptan, CH$_{3}$CH$_{2}$SH, in the millimeter and submillimeter-wave domains (up
to 880 GHz) provided very precise values of the spectroscopic constants that allowed the detection of $gauche$-CH$_3$CH$_2$SH
towards Orion KL. 77 unblended or slightly blended lines plus no missing transitions in the range 80~--~280~GHz support
this identification. A detection of methyl mercaptan, CH$_{3}$SH, in the spectral survey of Orion KL is reported as well.
Our column density results indicate that methyl mercaptan is $\simeq$\,5 times more abundant than ethyl mercaptan in the
hot core of Orion KL.
\end{abstract}


\section{Introduction}

The spectral millimeter-wave survey of Orion KL carried out with the IRAM 30m radio telescope presented more than
8000 unidentified lines \citep{Tercero2010, Tercero2012a}. Many of them (nearly 4000) have been identified as lines
arising from isotopologues and vibrationally excited states of abundant species (\citealt{Demyk2007, Carvajal2009,
Margules2009, Margules2010, Tercero2012b, Motiyenko2012, Daly2013, Coudert2013, Haykal2013b}, A. L\'{o}pez et al.
in preparation). These identifications significantly reduce the number of U-lines and mitigate line confusion in the spectra.
Nevertheless, many of those features still remain unidentified. Therefore, a search for
new molecular species in that cloud based on precise laboratory measurements continues to be a field of a great
research activity. Recently, the discovery of methyl acetate and \textit{gauche}-ethyl formate (\citealt{Tercero2013}),
the search for allyl isocyanide (\citealt{Haykal2013a}) as well as a tentative detection of phenol (\citealt{Kolesnikova2013})
have been reported.

Methanol and ethanol are very well known molecules in many astrophysical environments. The thiol equivalent of methanol,
methyl mercaptan (CH$_3$SH), has been detected towards Sgr\,B2 by \citet{Linke1979}. Previous searches for it towards
dark clouds and Orion did not provide positive results \citep{Irvine1987,Irvine1989}. However, this molecule has been
recently detected towards the cold first hydrostatic core B1 (\citealt{Cernicharo2012b}) and was also observed towards the
hot core G327.3-0.6 by \citet{Gibb2000}. 
Hence, the thiol equivalent of ethanol, ethyl mercaptan (CH$_3$CH$_2$SH), could also be present in space. Initial studies of the Stark modulated microwave spectra by \citet{Imanov67}, \citet{Hayashi71},
\citet{Hayashi73}, \citet{Quade75}, and \citet{Nakagawa76} provided the first values of the spectroscopic constants for the
\textit{gauche} and \textit{trans} conformers of ethyl mercaptan. These microwave laboratory data, however, cannot be used to accurately predict their frequencies in the millimeter and submillimeter-wave domains.

In this letter, we report new laboratory measurements of ethyl mercaptan in the millimeter and submillimeter-wave
region and its first observation in the interstellar medium. Newly derived spectroscopic constants of its \textit{gauche} and \textit{trans} forms have allowed us to
detect the $gauche$ conformer towards Orion KL with the IRAM 30m radio telescope. $Trans$-ethyl mercaptan has
been tentatively detected in this study. In addition, the first detection of methyl mercaptan towards Orion KL is
also presented.

\section{Laboratory Measurements}\label{sect-exp}
A commercial sample of ethyl mercaptan was used without any further purification. Rotational spectra in the
8.7~--~26.5~GHz region were taken at a temperature of $-15 ~\degr$C and a pressure of 6~mTorr with a
waveguide chirped-pulse Fourier transform microwave spectrometer at New College of Florida \citep{anisole}. The waveguide was cooled via a set of home-built cooling loops connected to a chiller with a recirculating fluid ($-15 ~\degr$C is the base temperature reached).
A 250 ns chirped pulse is generated with an arbitrary waveform generator (Tektronix AWG7101) and subsequently
frequency shifted, filtered, and amplified before interacting with a static gas sample in a 10m coil of WRD-750 waveguide.
The resulting molecular free induction decay is amplified and detected in the time-domain with an oscilloscope
(Tektronix TDS6154C), after which it is Fourier transformed to generate a frequency-domain spectrum. In each spectrum,
2 million free induction decays of 4~microsecond duration were averaged and Fourier transformed using a Kaiser-Bessel
window function. Peaks in the frequency-domain spectrum had a full width at half maximum of $\sim$~800~kHz and so
the uncertainty on the line center was set to 80~kHz.

Rotational spectra in the millimeter (60~--~300~GHz) and submilimeter-wave (500~--~590~GHz, 625~--~660~GHz, and
835~--~880~GHz) regions were recorded at a pressure of approximately 15~mTorr by a recently constructed millimeter-wave
spectrometer at the University of Valladolid (A. M. Daly et al. in preparation). The source of radiation is an Agilent E8257D synthesizer (250~kHz~ --~20~GHz) connected to a set of passive or active cascade frequency multipliers (VDI, Inc.). Room temperature measurements were
carried out in a free space 360~cm long Pyrex cell. Up to 170~GHz, the optical path length was doubled using a rooftop
mirror and a polarization grid. A silicon bolometer was used as the detection element
above 800~GHz. At lower frequencies, the signal was detected by either Schottky diodes or the Quasi-optical broadband detectors (VDI, Inc.). All the spectra were recorded in 1\,GHz sections using the frequency modulation technique with second harmonic lock-in detection (modulation
depth between 20~--~50~kHz and modulation frequency of 10.2~kHz for the semiconductor detectors and 90~Hz for the cryogenic detector).
Frequency accuracy is estimated to be better than 50~kHz. Rotational spectra of the $^{34}$S isotopologue from the microwave up to the submillimeter-wave region were measured in natural abundance.

\section{Rotational Spectra and Analysis}\label{sect-ana}

\begin{figure*}
\includegraphics[angle=0,scale=.7]{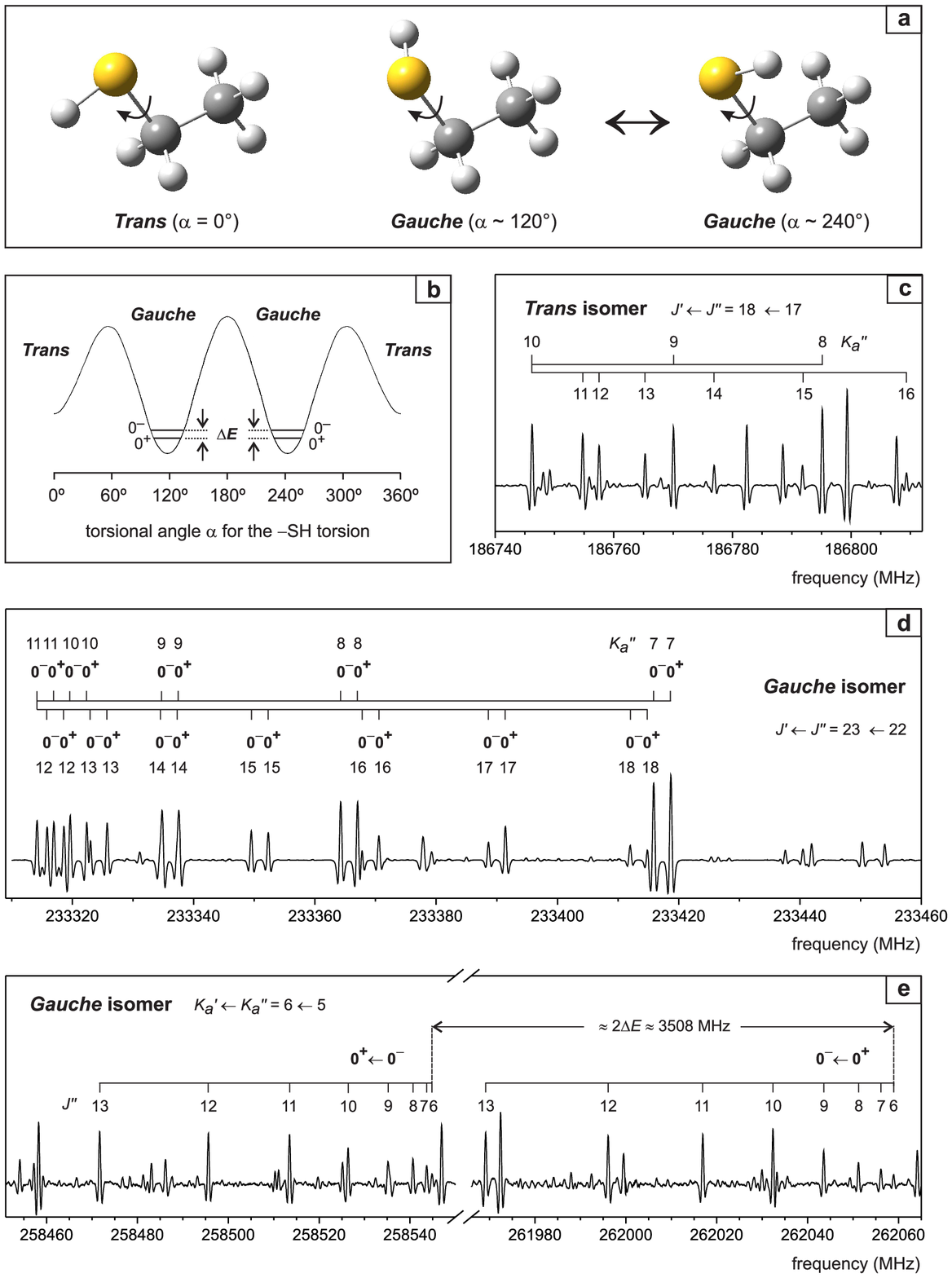}
\caption{Ethyl mercaptan: (a) \textit{trans} and two equivalent \textit{gauche} configurations; (b) potential energy curve for the --SH group torsion with torsion sublevels for the \textit{gauche} vibrational ground state; (c) $a$-type $R$-branch transitions of the \textit{trans} isomer; (d) pure rotational $a$-type $R$-branch transitions of the \textit{gauche} isomer; (e) torsion-rotational $c$-type $Q$-branch transitions of the \textit{gauche} isomer.} \label{exp}
\end{figure*}

Ethyl mercaptan is a near prolate asymmetric top molecule which is present in two stable forms: \textit{gauche} and
\textit{trans} configurations which are defined by the value of the
torsion angle $\alpha$ of the --SH group measured from the symmetry plane of the \textit{trans} configuration
($\alpha = 0$), see Figure \ref{exp}-a. For the \textit{gauche} conformer, two equivalent configurations at approximately
 $\alpha = 120\degr$ and $240\degr$ can be interchanged by tunneling motion. A schematic potential energy diagram for
the --SH group torsion is shown in Figure \ref{exp}-b. The tunneling process removes the vibrational degeneracy and the vibrational
ground state is split into two substates labeled as $0^{+}$ and $0^{-}$ with an energy separation $\Delta E = E^{-} - E^{+}$
of about 1750~MHz (\citealt{Quade75}, \citealt{Nakagawa76}).
In order to facilitate the assignments of the \textit{gauche}- and \textit{trans}-ethyl mercaptan millimeter-wave spectra,
the microwave data from \citet{Quade75} were used for initial predictions of the rotational transitions. Pickett's spectral
fitting programs SPCAT and SPFIT \citep{Pickett91} were used to predict the spectra as well as to determine the spectroscopic
constants. The visualization, processing and the assignments of the rotational spectra were performed using the SVIEW and ASCP
programs included in the AABS package \citep{aabs}.

The rotational spectrum of \textit{gauche}-ethyl mercaptan is formed by two types of transitions: $a$- and $b$-type
pure rotational transitions ($\mid\mu_{a}\mid = 1.48~(2)$~D,
$\mid\mu_{b}\mid = 0.19~(10)$~D \citep{Quade75}) between the rotational levels within the same torsion substate, i.e. $0^{\pm}\leftrightarrow
0^{\pm}$, and $c$-type torsion-rotational transitions ($\mid\mu_{c}\mid = 0.59~(2)$~D \citep{Quade75}) connecting different torsion substates ($0^{\mp} \leftrightarrow 0^{\pm}$).
Figure \ref{exp}-d illustrates a section of the $a$-type $R$-branch pure rotational transitions which are the dominant features
of the ethyl mercaptan rotational spectrum. The doubling pattern of the $c$-type transitions resulting from the tunneling process can
be seen in the Figure \ref{exp}-e. The frequency separation between the corresponding $c$-type doublets is close to $2\Delta E$,
which is about 3.5~GHz (see Figure \ref{exp}-e).

Several near degeneracies of the $0^{+}$ and $0^{-}$ energy levels beginning at low $J$ values have been observed which require accounting for the Coriolis-like terms that connect the $0^{+}$ and $0^{-}$ substates. Perturbation-allowed transitions between
$0^{+}$ and $0^{-}$ energy levels with $\Delta K_{a} =$ even and $\Delta K_{c} =$ even selection rules arising from the mixing of
the $0^{+}$ and $0^{-}$ substates with the same $K_{a}$ quantum number have been observed as well. Finally, more than 2200 distinct
transitions involving the rotational quantum numbers $J''$ and $K_{a}''$
up to 88 and 25, respectively, have been analyzed using the effective framework fixed axes two-state Hamiltonian as defined by
\citet{Quade63}:
\begin{equation}
H_{\mathrm{eff}} = H_{\mathrm{rot}}^{+} + H_{\mathrm{rot}}^{-} + H^{\pm} + \Delta E \label{Heff},
\end{equation}
where $H_{\mathrm{rot}}^{+}$ and $H_{\mathrm{rot}}^{-}$ represent the Watson's $A$-reduced semirigid Hamiltonian
in $I^{\mathrm{r}}$-representation \citep{Watson} for the $0^{+}$ and $0^{-}$ substates, respectively,
and $H^{\pm}$ is the torsion-rotation interaction Hamiltonian:
\begin{equation}
H^{\pm} = D^{\pm}(J_{b}J_{c}+J_{c}J_{b}) + Q^{\pm}J_{a} + N^{\pm}J_{b}\label{Htr},
\end{equation}
where $D^{\pm}$, $Q^{\pm}$, and $ N^{\pm}$ represent the Coriolis-like coupling constants. Centrifugal distortion corrections up
to the eighth order have been included in the analysis. All the perturbed transitions have been successfully treated by means of
the Coriolis-like constants defined in Equation \ref{Htr}, with the $K$-dependence of $D^{\pm}$ constant taken into account. For the
\textit{gauche}-$^{34}$S isotopologue, more than 700 transitions ($J''$ and $K_{a}''$ up to 56 and 18, respectively) have been
analyzed in the same manner as the parent species. The determined spectroscopic constants for both isotopologues are listed in
Table \ref{Tab1}. The measured frequencies are given in Table \ref{tab_ch3ch2sh} (the full Table \ref{tab_ch3ch2sh} is given in the Supplementary Material).

The rotational spectrum of \textit{trans}-ethyl mercaptan is dominated by pure rotational $a$- and $b$-type transitions ($\mid\mu_{a}\mid = 1.06~(3)$~D, $\mid\mu_{b}\mid = 1.17~(3)$~D \citep{Quade75}). An
example of measured $a$-type $R$-branch transitions is illustrated in Figure \ref{exp}-c. Many assigned \textit{trans} transitions
with higher values of $K_{a}''$ quantum numbers have been found to be affected by perturbations which may originate from
interactions with low-lying first excited states of the --SH and --CH$_{3}$ torsional modes. Only the unperturbed transitions
were included in the final fit. More than 600 distinct transitions with $J''$ and $K_{a}''$ quantum numbers up to 64 and 13,
respectively, were analyzed using the standard single-state semirigid $A$-reduced Hamiltonian \citep{Watson} with octic
centrifugal distortion constants. The resulting spectroscopic constants are given in Table \ref{Tab1}. The measured transitions
are given in the full Table \ref{tab_ch3ch2sh} (see the Supplementary Material).

\section{Astronomical Observations}\label{sect-obs}

Five observing sessions (from September 2004 to January 2007) were required for completing
a molecular line survey in all the frequencies available by the A, B, C, and D receivers (80~--~115.5, 130~--~178, 197~--~281 GHz)
of the IRAM 30m telescope towards Orion KL (IRc2 source at $\alpha$(J2000)=5$^{h}$ 35$^{m}$ 14.5$^{s}$,
$\delta$(J2000)=$-$5$\degr$ 22$\arcmin$ 30.0$\arcsec$).
System temperatures, image side band rejections, and half power beam widths were in the ranges 100~--~800 K, 27~--~13 dB,
and 29~--~9'', respectively, from lower to higher frequencies.
The intensity scale was calibrated using the ATM package \citep{Cernicharo1985, Pardo2001}.
The observations were performed in the balanced wobbler-switching mode.
As backends we connected two filter banks (512~MHz of bandwidth for each one and 1~MHz of spectral resolution) and a correlator (2$\times$512~MHz
of bandwidth and 1.25~MHz of spectral resolution).
Pointing and focus were checked every 1~--~2 hours on nearby quasars.
The data were processed with the GILDAS software\footnote{http://www.iram.fr/IRAMFR/GILDAS}.
The data reduction consisted of removing lines from the image side band and fitting and removing
baselines. Figures are shown in units of main beam temperature, T$_{MB}$=T$^{\star}_A$/$\eta_{MB}$, where $\eta_{MB}$ is the main beam efficiency.
More detailed description of the observations can be found in \cite{Tercero2010}.

At least four cloud components could be identified
in the line profiles of our low resolution spectral lines,
characterized by different radial velocities and line widths \citep{Blake1987,Schilke2001,Persson2007, Tercero2010, Tercero2011, Neill2013}.
Each component corresponds to a specific region of the cloud that overlaps in our telescope beam:
the extended ridge or ambient cloud (T$_K$ $\simeq$ 60 K); the compact ridge, a dense clump characterized by
the emission of organic saturated O-rich molecules at a T$_K$ $\simeq$ 150 K; the plateau, or outflow from
the new born stars (T$_K$ $\simeq$ 150 K); and the hot core, a dense and warm region rich in organic saturated N-bearing
molecules (T$_K$ $\simeq$ 250 K).

\section{Results and Discussion}\label{sect-det}

\begin{figure*}
\includegraphics[angle=0,scale=.7]{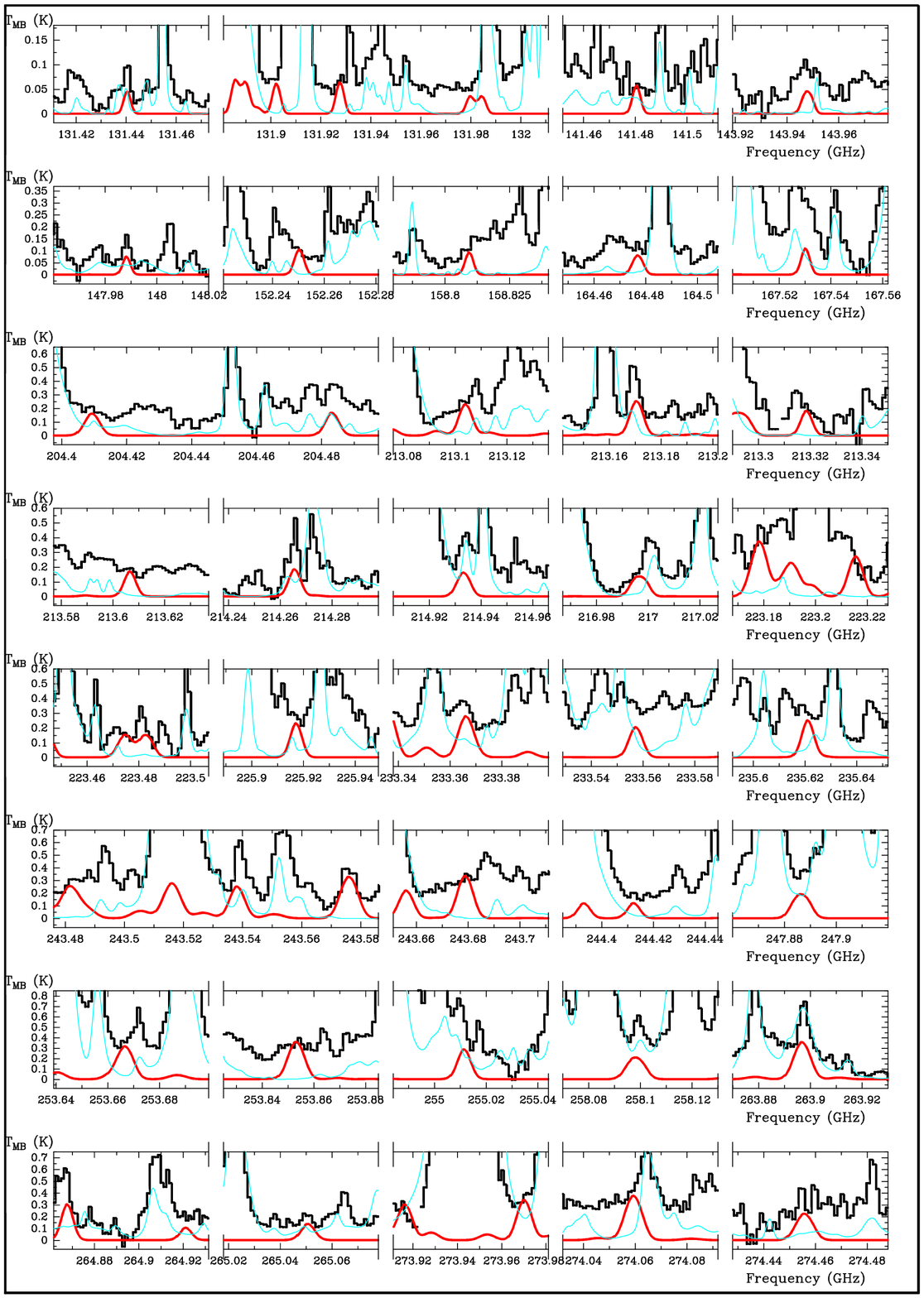}
\caption{Selected lines of $gauche$-ethyl mercaptan
towards Orion-IRc2 (in red). The continuous cyan line
corresponds to all lines already modeled in our previous papers (see text). A $v$$_{LSR}$ of 5\,km\,s$^{-1}$ is assumed.} \label{fig-ethyl}
\end{figure*}

Direct laboratory measurements and derived spectroscopic constants given in Table \ref{Tab1} have allowed us to detect $gauche$-ethyl mercaptan
and to tentatively detect the $trans$ conformer in the molecular line survey of Orion KL by means of a large number of spectral lines free
of blending with other species. Table \ref{tab_ch3ch2sh} provides (together with the transitions measured
in the laboratory) the observational parameters of the detected lines that are not strongly blended with other molecules. Owing to the weakness
of these features, the main beam temperature has been obtained from the peak channel of the spectra, so errors in the baselines and contribution
from other species could affect this parameter. Therefore, T$_{MB}$ has to be considered as the total intensity of the detected feature and
an upper limit to the intensity of ethyl mercaptan in this study.
A total amount of 34 transitions free of blending (in the peak channel of the feature)
and 43 slightly blended lines are reported in Table \ref{tab_ch3ch2sh} for $gauche$--ethyl mercaptan. No missing lines have been found for this
conformer. Figure \ref{fig-ethyl} shows some of the lines of $gauche$-ethyl mercaptan reported here.
Even though the $gauche$ conformer being more stable, the number of potential lines to be detected for the
$trans$ conformer could be larger due to the higher value of its $b$-dipole moment component (see above).
For $trans$-ethyl mercaptan, 72 lines free of blending and 52 slightly blended lines could be present in the survey. They are listed
in Table \ref{tab_ch3ch2sh} with the estimate of their parameters. However, the weakness of the transitions of $trans$-CH$_3$CH$_2$SH
and the high overlap with other molecules make difficult the assignment of isolated lines of this conformer as they appear at the
confusion limit of the line survey. Hence, we consider only a tentative
detection for the $trans$ conformer of ethyl mercaptan (see below).
We also searched for CH$_3$CH$_2$SH in the PRIMOS survey\footnote{Hollis, J. M.; Remijan, A. J.; Jewell, P. R.; Lovas, F. J.; Corby, J. F.
http://www.cv.nrao.edu/$\sim$aremijan/PRIMOS/index.html} with the Green Bank Telescope (GBT) in the frequency range between 300\,MHz and 50\,GHz and in the
$>$13$\sigma$ U lines of the 3\,mm IRAM 30m survey reported by \citet{Belloche2013}, both surveys towards SgrB2, finding a negative detection in both set of data.

\begin{figure*}
\includegraphics[angle=0,scale=.7]{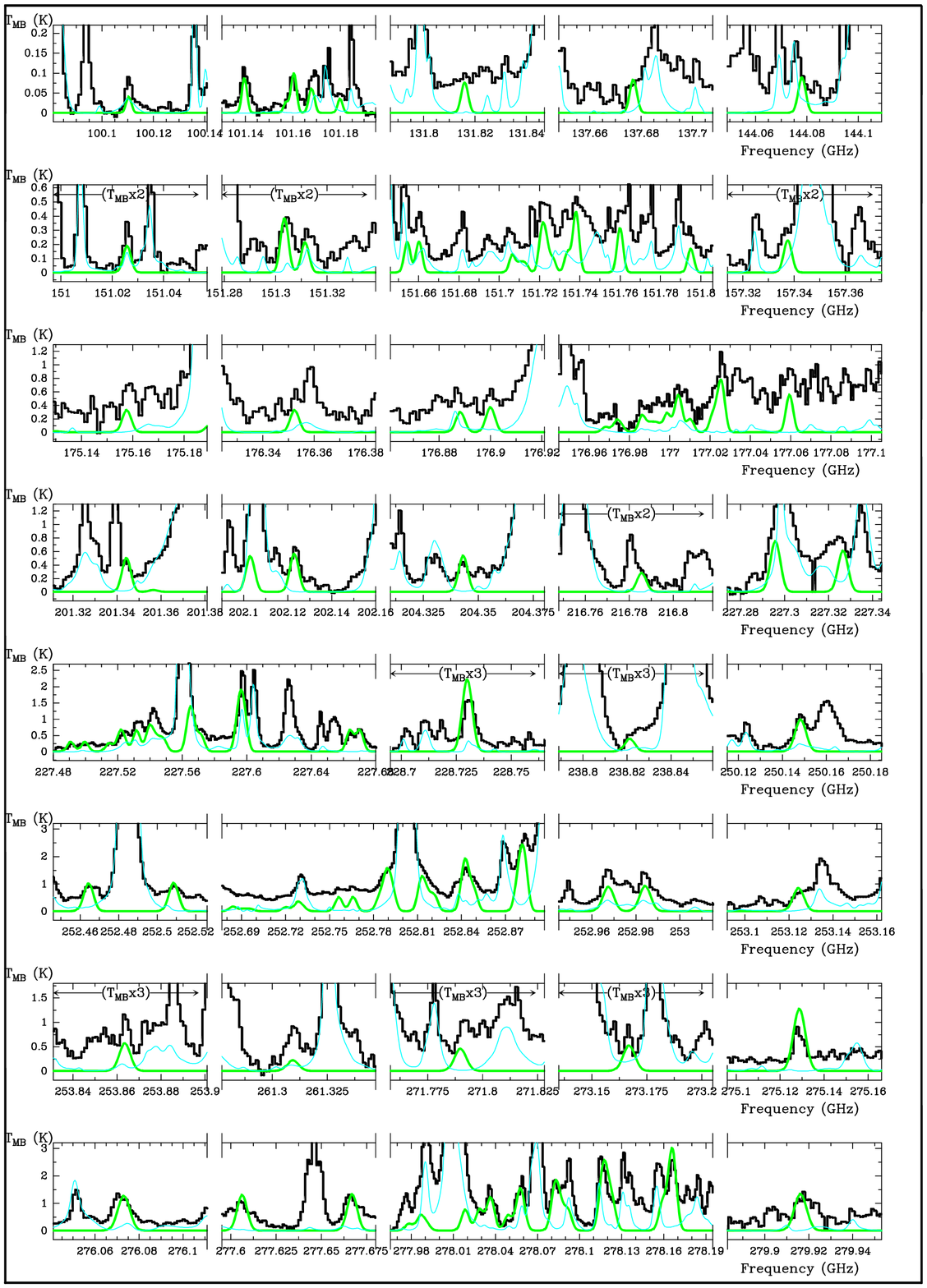}
\caption{Selected lines of methyl mercaptan towards Orion-IRc2. The continuous cyan line
corresponds to all lines already modeled in our previous papers (see text). A $v$$_{LSR}$ of 5\,km\,s$^{-1}$ is assumed.} \label{fig-methyl}
\end{figure*}

In order to compare the emission of ethyl mercaptan with that of methyl mercaptan, the latter species was searched for in Orion KL.
The laboratory frequencies and dipole moment values used for the methyl mercaptan are those predicted by \citet{Bettens1999} and
have been implemented in the MADEX code \citep{Cernicharo2012a}.
Figures \ref{fig-ethyl} and \ref{fig-methyl} show selected lines of $gauche$-CH$_3$CH$_2$SH and A/E-CH$_3$SH, respectively.
Both Figures show our best model for these species (see below),
and the total model for the already studied species in this survey (see \citealt{Tercero2013} and references therein).
The figures show lines free of blending, or moderately blended with lines from other species. No missing lines have been found
for methyl mercaptan in the frequency range covered by our line survey.

To model the emission of A/E-CH$_3$SH and $gauche$/$trans$-CH$_3$CH$_2$SH we have considered
that both molecules come from the same
region of Orion KL (hot core at $v_{LSR}$\,=\,5\,km\,s$^{-1}$ and $\Delta$$v$\,=\,7\,km\,s$^{-1}$).
Lines from methyl mercaptan are typically 10 times stronger than those of ethyl mercaptan.
We used the MADEX code in LTE conditions due to the lack of
collisional rates for these species.
Beam dilution and the position of the source with respect to the pointing position were
taken into account in our models. We assumed a $d_{sou}$\,=\,10'' and $offset$\,=\,3''. In \cite{Tercero2010}
we estimated the uncertainties of the column density results in this survey to be between 20~--~30 \% considering
different sources of uncertainty such us the spatial overlap of the different cloud components, the
modest angular resolution of any single-dish line survey or
pointing errors. Nevertheless, the uncertainty due to high overlap problems has to be considered for
results obtained by means of weak lines such as those of ethyl mercaptan (raising the uncertainty up to 50 \%).
The physical and chemical parameters derived by the model are a common kinetic
temperature of 200$\pm$50\,K for ethyl and methyl mercaptan and a column density of
(5.0$\pm$2.0)$\times$10$^{15}$\,cm$^{-2}$ for each A and E state of CH$_3$SH and of (2.0$\pm$1.0)$\times$10$^{15}$\,cm$^{-2}$
for $gauche$-ethyl mercaptan. For $trans$-CH$_3$CH$_2$SH we can report only a tentative detection with an upper limit to its column density
of $\leq$(2.0$\pm$1.0)$\times$10$^{15}$\,cm$^{-2}$. Therefore,
we found that methyl mercaptan is $\simeq$5 times more abundant than ethyl mercaptan in the hot core of Orion KL.
The population of different conformers of CH$_3$CH$_2$SH follows a Boltzmann distribution so the abundance ratio between both
conformers of ethyl mercaptan is given by $N(trans)$/$N(gauche)$\,=\,exp(-$E_{rel}$/$T_k$), where $E_{rel}$
is the difference of energy between the conformers (230\,K, M. L. Senent, private communication) and $T_k$ is the
kinetic temperature of the medium. At 200\,K this formula yields a $N(trans)$/$N(gauche)$ value of 0.32 whereas at
300\,K this number rises up to 0.46. These abundance ratios
have to be corrected by the vibrational partition function of both conformers.
For T$_K$=200\,K the vibrational partition function for $gauche$ is 1.47 and for $trans$ it is 1.54 (M. L. Senent, private communication).
Applying these corrections, our abundance ratio between $trans$ and $gauche$ (both states 0$^+$ and 0$^-$) is $\leq$1.

In order to compare these results with structurally similar molecules, we modeled the emission of ethanol ($trans$/$gauche$-CH$_3$CH$_2$OH)
and methanol (A/E-$^{13}$CH$_3$OH) in our line survey using the MADEX code and LTE conditions (A. L\'opez et al., in preparation). For
fitting the line profiles in this large spectral range, five cloud components are required: the four components described
in Sect.\,\ref{sect-obs} and a hotter (300\,K) and smaller ($d_{sou}$=7'') compact ridge. For the four former components,
the assumed physical conditions are those derived in \cite{Tercero2010}. We obtained a column densities of (6$\pm$2)$\times$10$^{17}$\,cm$^{-2}$,
(2.2$\pm$0.6)$\times$10$^{16}$\,cm$^{-2}$, and (1.7$\pm$0.4)$\times$10$^{16}$\,cm$^{-2}$ for each state of CH$_3$OH (assuming a
$^{12}$C/$^{13}$C ratio of 45, see \citealt{Tercero2010}), for the $trans$ conformer of ethanol, and for $gauche$-CH$_3$CH$_2$OH,
respectively. Therefore, methanol is 30 times more abundant than ethanol in Orion KL. This difference between $X$(CH$_3$OH/CH$_3$CH$_2$OH)
and $X$(CH$_3$SH/CH$_3$CH$_2$SH) could be due to the methanol and methyl mercaptan stay time on the dust grains before evaporation. If
methyl mercaptan stays on the dust grains longer time than methanol, then, further chemical processing could occur changing the
abundance ratios between methyl and ethyl species. Taking into account that the emission of the $-$OH species comes mainly from the
compact ridge whereas methyl and ethyl mercaptan emit from the hot core, the differences of relative abundances between the methyl
and ethyl species could be also due to the chemical differentiation between these two regions inside Orion KL.
Most sensitive observations, with interferometers such us ALMA are needed to derive accurate column densities
of the two conformers of ethyl mercaptan and to study their spatial distribution in Orion.

\clearpage

\acknowledgments

The authors thank the Spanish MINECO for support under grants AYA 2009-07304, AYA2011-29375,
CTQ 2006-05981/BQU, CTQ 2010-19008, and the CONSOLIDER program "ASTROMOL" CSD 2009-00038 and Junta
de Castilla y Leon (Grant VA070A08). B.P.G. and S.T.S. acknowledge support from the National Science
Foundation Division of Chemistry under Grant No. 1111101 (co-funded by MPS/CHE and the Division of
Astronomical Sciences) and B.P.G. also thanks New College of Florida for a Student Research and
Travel Grant award.

\clearpage




\clearpage

\begin{deluxetable}{llrrrrr}
\tabletypesize{\scriptsize}
\tablecolumns{7}
\tablewidth{0pt}
\tablecaption{Spectroscopic constants\tablenotemark{a} of \textit{gauche}- and \textit{trans}-ethyl mercaptan ($A$-reduction, $I^{\mathrm{r}}$-representation).\label{Tab1}}
\tablehead{
& & \multicolumn{2}{c}{\textit{Gauche} (parent specie)}& \multicolumn{2}{c}{\textit{Gauche} ($^{34}$S)}\\
\cline{3-6}
\colhead{Constant} & \colhead{Unit} & \colhead{$0^{+}$} & \colhead{$0^{-}$}  &\colhead{$0^{+}$} & \colhead{$0^{-}$} & \colhead{\textit{Trans}}
}
\startdata
$A$                      &(MHz) &28747.4104 (66)   &28747.2715 (65)  & 28709.699 (24)                    &28708.964 (26)               &28416.7604 (18)\\
$B$                      &(MHz) &5295.1422 (36)    &5295.0008 (36)   &  5175.5685 (36)                   & 5175.4383 (36)              &5485.77901 (15)\\
$C$                      &(MHz) &4845.9421 (36)    &4845.9689 (36)   &  4744.7050 (36)                   & 4744.7310 (36)              &4881.81770 (15)\\
$\Delta_{J}$             &(kHz) &3.326369 (20)     &3.323582 (20)    &     3.18351 (18)                  &    3.18084 (18)             &3.83217 (27) \\
$\Delta_{JK}$            &(kHz) &$-$18.39280 (60)  &$-$18.35859 (60) & $-$17.9512 (15)                   &  -17.8994 (14)              &$-$22.4549 (58)\\
$\Delta_{K}$             &(kHz) &204.1591 (82)     &203.9217 (81)    &   206.8 (14)                      &  205.1 (14)                 &210.03 (18) \\
$\delta_{J}$             &(kHz) &0.514429 (12)     &0.513170 (14)    &     0.481132 (82)                 &    0.480129 (71)            &0.65664 (11) \\
$\delta_{K}$             &(kHz) &8.781 (12)        &8.555 (12)       &     8.271 (34)                    &    8.024 (35)               &7.342 (20)\\
$\Phi_{J}$               &(kHz) &0.0029872 (28)    &0.0029225 (31)   & 0.0029872\tablenotemark{b}        & 0.0029225\tablenotemark{b}  &$-$0.00086 (17)\\
$\Phi_{JK}$              &(kHz) &0.0790 (13)       &0.0661 (13)      & 0.0790\tablenotemark{b}           & 0.0661\tablenotemark{b}     &0.367 (18) \\
$\Phi_{KJ}$              &(kHz) &$-$1.3341 (45)    &$-$1.2955 (44)   & $-$1.3341\tablenotemark{b}        & $-$1.2955\tablenotemark{b}  &$-$2.717 (106) \\
$\Phi_{K}$               &(kHz) &6.341 (48)        &5.171 (46)       & 6.341\tablenotemark{b}            & 5.171\tablenotemark{b}      &26.7 (80)\\
$\phi_{J}$               &(kHz) &0.0012107 (15)    &0.0011805 (16)   & 0.0012107\tablenotemark{b}        & 0.0011805\tablenotemark{b}  &$-$0.000935 (91) \\
$\phi_{JK}$              &(kHz) &0.04561 (63)      &0.04326 (65)     & 0.04561\tablenotemark{b}          & 0.04326\tablenotemark{b}    &$-$0.164 (26) \\
$\phi_{K}$               &(kHz) &5.858 (95)        &4.825 (97)       & 5.858\tablenotemark{b}            & 4.825\tablenotemark{b}      &28.89 (97) \\
$L_{J}$                  &(mHz) & ...              &...              & ...                               & ...                         &0.000060 (11) \\
$L_{KKJ}$                &(mHz) & ...              &...              & ...                               & ...                         &2.11 (40)\\
$L_{K}$                  &(mHz) &$-$1.140 (75)     &0.673 (71)       & ...                               & ...                         &1374 (110) \\
$l_{JK}$                 &(mHz) & ...              &...              & ...                               & ...                         & 0.0497 (61) \\
$l_{KJ}$                 &(mHz) &0.0523 (38)       &0.0414 (38)      & ...                               & ...                         & 4.77(50) \\
$l_{K}$                  &(mHz) &0.403 (21)        &0.601 (19)       & ...                               & ...                         &...  \\
$\Delta E$               &(MHz) &\multicolumn{2}{c}{1753.9788 (37)}  &\multicolumn{2}{c}{1730.995 (19)}                                &...  \\
$D^{\pm}$                &(MHz) &\multicolumn{2}{c}{8.359 (96)}      &\multicolumn{2}{c}{5.36 (14)}                                    &...  \\
$Q^{\pm}$                &(MHz) &\multicolumn{2}{c}{$-$15.00 (38)}   &\multicolumn{2}{c}{$-$26.51 (57)}                                &...  \\
$N^{\pm}$                &(MHz) &\multicolumn{2}{c}{6.30 (11)}       & ...                               & ...                         &...  \\
$D^{\pm}_{K}$            &(MHz) &\multicolumn{2}{c}{0.01358 (50)}    & ...                               & ...                         &...  \\
$\sigma_{\mathrm{fit}}$  &(kHz) &\multicolumn{2}{c}{39}              &\multicolumn{2}{c}{41}                                           & 30\\
\enddata
\tablenotetext{a}{The numbers in parentheses are 1$\sigma$ uncertainties in the units of the last decimal digit.}
\tablenotetext{b}{Fixed to the parent species value.}
\end{deluxetable}

\clearpage

\begin{deluxetable}{rrrrrrrrrrrrrrrl}
\tablewidth{0pt}
\tabletypesize{\scriptsize}
\tablecolumns{16}
\tablecaption{Laboratory measurements and astronomical detected lines of ethyl mercaptan.\label{tab_ch3ch2sh}}
\tablehead{
\colhead{$J'$}           & \colhead{$K_{a}'$}      &
\colhead{$K_{c}'$}       & \colhead{$v'$}  &
\colhead{$J''$}          & \colhead{$K_{a}''$}      &
\colhead{$K_{c}''$}      & \colhead{$v''$}  &
\colhead{Laboratory}     & \colhead{Lab.$-$Pred.} &
\colhead{Conf.}          & \colhead{$E_u$}      &
\colhead{$S_{ij}$}       & \colhead{Sky Obs. freq.}      &
\colhead{$T_{mb}$}       & \colhead{Blend}\\

\colhead{}               & \colhead{}      &
\colhead{}               & \colhead{}  &
\colhead{}               & \colhead{}      &
\colhead{}               & \colhead{}  &
\colhead{freq. (MHz)}    & \colhead{(MHz)} &
\colhead{}               & \colhead{(K)}      &
\colhead{}               & \colhead{(MHz)}      &
\colhead{(K)}            & \colhead{}}

\startdata
 13 &  5 &  8 &  1 & 12 &  5 &  7 &  1 &  131926.7036     & -0.0205  &       G&  72.8 &  11.1 &   131928.9  & 0.08 &                                                               \\
 13 &  5 &  9 &  1 & 12 &  5 &  8 &  1 &  131926.7036     &  0.0443  &       G&  72.8 &  11.1 &   $\dagger$ &      &                                                               \\
 13 &  5 &  8 &  0 & 12 &  5 &  7 &  0 &  131928.2800     & -0.0229  &       G&  72.7 &  11.1 &   $\dagger$ &      &                                                               \\
 13 &  5 &  9 &  0 & 12 &  5 &  8 &  0 &  131928.2800     &  0.0420  &       G&  72.7 &  11.1 &   $\dagger$ &      &                                                               \\
 13 &  4 & 10 &  1 & 12 &  4 &  9 &  1 &  131978.8083     & -0.0052  &       G&  62.6 &  11.8 &   131977.9  & 0.07 &                                                               \\
 13 &  4 & 10 &  0 & 12 &  4 &  9 &  0 &  131980.4297     & -0.0062  &       G&  62.5 &  11.8 &   $\dagger$ &      &                                                               \\
 13 &  4 &  9 &  1 & 12 &  4 &  8 &  1 &  131983.4352     & -0.0123  &       G&       &       &             &      &                                                               \\
 13 &  4 &  9 &  0 & 12 &  4 &  8 &  0 &  131985.0626     & -0.0126  &       G&       &       &             &      &                                                               \\
 13 &  3 & 11 &  1 & 12 &  3 & 10 &  1 &  132035.3178     & -0.0038  &       G&  54.6 &  12.3 &   132033.3  & 0.03 &  \tiny{CH$_3$COOCH$_3$}                                       \\
\enddata
\tablecomments{This table is published in its entirety in the
electronic edition of the {\it Astrophysical Journal Letters}.  A portion is
shown here for guidance regarding its form and content.
Table gives the laboratory measurements and emission lines of the \textit{gauche}- and \textit{trans}-CH$_3$CH$_2$SH
present in the Orion KL survey. All the transition frequencies included in the least-square analysis were weighted inversely proportional to the second power of their estimated uncertainties. The unresolved pairs of the asymmetry splitting were always included as two overlapped transitions and fitted to their intensity weighted averages. Columns 1--8 indicate the upper and lower state quantum numbers. Col. 9
gives the measured frequency at the laboratory. Col. 10 gives the difference between
frequency measurements and predictions. Col. 11 gives the corresponding conformer.
Col. 12 gives the upper level energy. Col. 13 gives the line strength.
Col. 14 gives the observed frequency in the line survey of Orion assuming a $v_{LSR}$ of 5 km\,s$^{-1}$. Col. 15 gives the main beam temperature. Col. 16 indicates blended transitions with other molecules.
$\dagger$ Blended with previous line.}
\end{deluxetable}

\end{document}